\documentclass[english,a4paper,preprintnumbers,pre,twocolumn]{revtex4-1}
\usepackage[T1]{fontenc}
\usepackage[latin9]{inputenc}
\usepackage[a4paper]{geometry}
\usepackage{amsmath}
\usepackage{amssymb}
\usepackage{graphicx}
\usepackage{esint}

\makeatletter
\@ifundefined{textcolor}{}
{%
 \definecolor{BLACK}{gray}{0}
 \definecolor{WHITE}{gray}{1}
 \definecolor{RED}{rgb}{1,0,0}
 \definecolor{GREEN}{rgb}{0,1,0}
 \definecolor{BLUE}{rgb}{0,0,1}
 \definecolor{CYAN}{cmyk}{1,0,0,0}
 \definecolor{MAGENTA}{cmyk}{0,1,0,0}
 \definecolor{YELLOW}{cmyk}{0,0,1,0}
 }

\usepackage{graphicx}

\makeatother

\usepackage{babel}
\begin{document}

\title{Entropy production from stochastic dynamics in discrete full phase
space}

\author{Ian J. Ford and Richard E. Spinney }

\address{Department of Physics and Astronomy, University College London, Gower
Street, London WC1E 6BT, United Kingdom and London Centre for Nanotechnology,
17-19 Gordon Street, London WC1H 0AH, United Kingdom}
\begin{abstract}
The stochastic entropy generated during the evolution of a system
interacting with an environment may be separated into three components,
but only two of these have a non-negative mean. The third component
of entropy production is associated with the relaxation of the system
probability distribution towards a stationary state and with nonequilibrium
constraints within the dynamics that break detailed balance. It exists
when at least some of the coordinates of the system phase space change
sign under time reversal, and when the stationary state is asymmetric
in these coordinates. We illustrate the various components of entropy
production, both in detail for particular trajectories and in the
mean, using simple systems defined on a discrete phase space of spatial
and velocity coordinates. These models capture features of the drift
and diffusion of a particle in a physical system, including the processes
of injection and removal and the effect of a temperature gradient.
The examples demonstrate how entropy production in stochastic thermodynamics
depends on the detail that is included in a model of the dynamics
of a process. Entropy production from such a perspective is a measure
of the failure of such models to meet Loschmidt's expectation of dynamic
reversibility.
\end{abstract}
\maketitle

\section{Introduction}

Entropy production is a measure of the irreversibility of a thermodynamic
process: the difficulty, even impossibility of reversing the observed,
often macroscopic behaviour of a system that exchanges heat or matter
with a complex environment. The apparent breakage of time reversal
symmetry associated with thermodynamic irreversibility has attracted
discussion for more than a century. It was given a particular focus
by Loschmidt's views on Boltzmann's work in gas dynamics \cite{Cercignani98}
but the issue was apparent far earlier in the contrast between Fourier's
law of heat flow and Newton's laws in mechanics. But in spite of such
concerns, the concept of entropy generation in the thermodynamics
of large systems has been developed and applied widely, to the extent
that equations for macroscopic entropy generation and transport associated
with nonequilibrium hydrodynamic and thermal flows are available,
for example see \cite{ChaikinLubensky95,Balescu97,EvansMorriss08}.

However, a microscopic understanding of the nature of entropy and
its production has proved elusive, particularly with regard to understanding
the one-way character of the second law. But some considerable steps
forward have been made by modelling the microscopic evolution of a
system and its environment using a framework of stochastic dynamics
\cite{Gardiner09} and stochastic thermodynamics \cite{seifertoriginal,seifertprinciples,sekimoto2}.
Such an approach explicitly breaks time reversal symmetry through
the use of a simplified model of the interactions between system and
environment, a consequence of coarse-graining, such that a second
law can emerge naturally. Nevertheless, it is imperative to define
entropy in this framework in such a way that makes contact with results
known to hold macroscopically. Entropy is often interpreted as a measure
of uncertainty at the microscopic level, and the viewpoint offered
by stochastic thermodynamics is that its overall production is a reflection,
according to some views verging on the tautological \cite{shargel},
of the uncertain or stochastic dynamics that drive the evolution,
though other interpretations based on deterministic dynamics exist
as well \cite{Evans94,Gallavotti2,Evans2011}.

The stochastic approach suggests that a change in thermodynamic entropy,
satisfying the second law, is an expectation value, or more simply
the mean, of the change in a microscopic, path-dependent entropy that
evolves stochastically in line with the dynamical model of the evolution
of the system \cite{seifertoriginal,seifertprinciples,sekimoto2}.
This view allows us to consider entropy as an instantaneous microscopic
property of a system and its environment, based on assignments of
probabilities to each of the available microstates, but with the potential
to increase or decrease during the process in question. This confers
meaning to the entropy change that follows from a single realisation
of a process, and avoids associating entropy production only with
some sort of average over many realisations. It is only after such
an averaging procedure that the total entropy is expected to increase.
Thus realisations that reduce entropy are quite possible: the second
law within this framework quite clearly holds only in a statistical
sense.

In detail, the entropy change associated with a specific evolution
of a system, referred to as a trajectory or path, is defined in stochastic
thermodynamics in terms of the probability that the path might be
realised under the prevailing `forward' dynamics, driven by a forward
protocol of time dependence, and the probability of realisation of
a path that represents the reversal of the first, under the same dynamical
rules but driven by a reversed protocol \cite{kurchan,GCforstochastic,Jarpathintegral,adiabaticnonadiabatic0,Harris07}.
This definition strongly associates the concept of entropy change
with that of dynamical irreversibility. When the system state is described
in a full phase space of spatial and velocity variables, or indeed
any set that includes variables that change sign under time reversal,
the reversed path clearly corresponds to a set of points in phase
space that retraces the sequence of spatial positions, but with velocity
coordinates that are inverted. It should be noted that the sequence
where the velocities are not reversed is sometimes inaccessible under
the given dynamics. The probability of the starting configuration
for the reversed path is specified to be that which is generated from
the forward process. Such a specification produces a path-dependent
entropy change that satisfies a number of requirements (see, for example,
\cite{SpinneyFordChapter12}).

These issues have recently been explored in the context of a master
equation describing the evolution of system probabilities over a discrete
full phase space \cite{SpinneyFord12a}, and for the stochastic dynamics
of continuous coordinates that transform differently under time reversal
\cite{SpinneyFord12b}. It has been established that the total path-dependent
entropy change for such cases may be divided into three components,
only two of which, together with the sum of the three, are expected
to be non-negative on average. In this paper we analyse a number of
examples involving a discrete full phase space, in order to provide
a greater appreciation of this division of entropy production, and
to reflect on the meaning of entropy production when the dynamics
of a system and its environment are considered at various levels of
detail under coarse-graining. We begin with a summary of the division
of path-dependent entropy production into its components, before discussing
particle drift and diffusion, flow brought about by injection and
removal, and the transport of heat brought about through interaction
of the particle with a temperature gradient. We end with some conclusions
and remarks on how entropy production, defined within a framework
of stochastic thermodynamics, in essence quantifies the failure of
Loschmidt's expectation of time reversal symmetry and dynamical reversibility
for a given set of circumstances.

\section{Formalism of entropy production}

Consider a particle that can occupy one of $L$ discrete spatial positions
$X_{i}$ in one dimension, and assume one of $M$ discrete values
of velocity $V_{m}$. The particle can make stochastic transitions
between phase space points $(X_{i},V_{m})\equiv(i,m)$. We define
the forward dynamics in terms of a transition rate $T(i^{\prime},m^{\prime}\vert i,m)$
characterising a move from $(i,m)$ to $(i^{\prime},m^{\prime})$,
such that the probabilities of occupation of the phase space points
$P(i,m,t)$ evolve according to the Markovian master equation
\begin{equation}
\frac{dP(i,m,t)}{dt}=\sum_{i^{\prime},m^{\prime}}T(i,m\vert i^{\prime},m^{\prime})P(i^{\prime},m^{\prime},t),\label{eq:1a}
\end{equation}
where the notation $T(i,m\vert i,m)$ represents the total rate of
transitions from point $(i,m)$:
\begin{equation}
T(i,m\vert i,m)=-\sum_{i^{\prime}\ne i,m^{\prime}\ne m}T(i^{\prime},m^{\prime}\vert i,m).\label{eq:2a}
\end{equation}
These dynamical rules can be employed to generate stochastic paths
of a particle across the discrete phase space. Master equation treatments
of dynamics in a discrete phase space of velocities as well as positions
have a long history \cite{vankampen76,brenig80}.

\begin{figure}
\includegraphics[width=1\columnwidth]{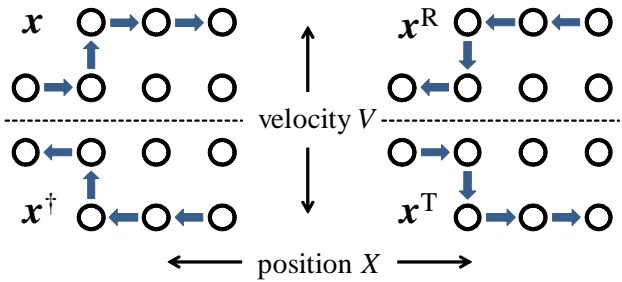}

\caption{The phase space coordinate sequence $\boldsymbol{x}$, with its associated
set of residence times, and its time-reversed counterpart $\boldsymbol{x}^{\dagger}$
, both shown on the left, can be generated by the forward dynamics.
The probabilities of each are used to define the total entropy production
$\Delta\mathcal{S}_{{\rm tot}}$ of the path $\boldsymbol{x}$. The
probabilities of a further two paths $\boldsymbol{x}^{R}$ and $\boldsymbol{x}^{T}$,
shown on the right on the same grid, are considered according to adjoint
dynamics, and serve to define contributions $\Delta\mathcal{S}_{1}$
and $\Delta\mathcal{S}_{2}$.\label{fig:Trajectories-in-phase}}
\end{figure}

Stochastic entropy production is associated with the probabilities
of dynamically generating a path and its reverse, and in a phase space
with coordinates that change sign under time reversal, such as a velocity,
the reversal of a path naturally requires an inversion of those coordinates.
This is indicated in Fig. \ref{fig:Trajectories-in-phase}, where
we denote a path by the sequence of phase space points $\boldsymbol{x}\equiv\left\{ \boldsymbol{x}_{0},\cdots,\boldsymbol{x}{}_{k},\cdots,\boldsymbol{x}_{N}\right\} $,
each point represented by the coordinates $(X_{i},V_{m})$. A particle
resides at point $\boldsymbol{x}_{k}$ in the time interval $t_{k}\le t\le t_{k+1}$
such that the duration of the path is from $t_{0}$ to $t_{N+1}$.
The reversed sequence of spatial coordinates and inverted velocity
coordinates, visited for a reversed sequence of intervals and denoted
$\boldsymbol{x}^{\dagger}\equiv\left\{ \boldsymbol{\varepsilon}\boldsymbol{x}_{N},\cdots,\boldsymbol{\varepsilon}\boldsymbol{x}{}_{k},\cdots,\boldsymbol{\varepsilon}\boldsymbol{x}_{0}\right\} $,
where $\boldsymbol{\varepsilon}$ indicates the change in sign of
the velocity coordinate (such that $\boldsymbol{\varepsilon}X_{i}=X_{i}$
and $\boldsymbol{\varepsilon}V_{m}=-V_{m})$, is the appropriate path
to use in defining the total entropy change associated with path $\boldsymbol{x}$:
\begin{equation}
\Delta\mathcal{S}_{{\rm tot}}(\boldsymbol{x})=\ln\left(P^{{\rm F}}(\boldsymbol{x})/P^{{\rm R}}(\boldsymbol{x}^{\dagger})\right),\label{eq:5}
\end{equation}
in units of the Boltzmann constant, where $P^{{\rm F}}(\boldsymbol{x})$
and $P^{{\rm R}}(\boldsymbol{x}^{\dagger})$ are the probabilities
that the sequences $\boldsymbol{x}$ and $\boldsymbol{x}^{\dagger}$
are generated. Explicitly, the path probability $P^{{\rm F}}(\boldsymbol{x})$
may be written
\begin{equation}
\begin{alignedat}{1} & P^{{\rm F}}(\boldsymbol{x})=e^{\int_{t_{N}}^{t_{N+1}}\! T(\boldsymbol{x}_{N}\vert\boldsymbol{x}_{N})dt^{\prime}}T(\boldsymbol{x}_{N}\vert\boldsymbol{x}_{N-1})dt_{N}\times\\
 & e^{\int_{t_{N-1}}^{t_{N}}\! T(\boldsymbol{x}_{N-1}\vert\boldsymbol{x}_{N-1})dt^{\prime}}T(\boldsymbol{x}_{N-1}\vert\boldsymbol{x}_{N-2})dt_{N-1}\times\\
 & \cdots T(\boldsymbol{x}_{1}\vert\boldsymbol{x}_{0})dt_{1}e^{\int_{t_{0}}^{t_{1}}\! T(\boldsymbol{x}_{0}\vert\boldsymbol{x}_{0})dt^{\prime}}\! P(\boldsymbol{x}_{0},t_{0}),
\end{alignedat}
\label{eq:5a}
\end{equation}
where $P(\boldsymbol{x}_{0},t_{0})$ is the probability that the system
resides at $\boldsymbol{x}_{0}$ at the initial time $t_{0}$, and
$T(\boldsymbol{x}_{j+1}|\boldsymbol{x}_{j})$ is synonymous with $T(i_{j+1},m_{j+1}\vert i_{j},m_{j})$.
The path probability is a product of component probabilities of residence
and transition. $P^{{\rm R}}(\boldsymbol{x}^{\dagger})$ is constructed
similarly, but the initial distribution is specified to be that which
generated by all possible paths corresponding to the forward process,
namely the solution to Eq. (\ref{eq:1a}) at $t=t_{N+1}$ starting
from a distribution corresponding to $P(\boldsymbol{x}_{0},t_{0})$.
The superscripts refer to the time dependence of the transition rates:
a forward protocol of time dependence is signified by an F, and R
denotes the reversal of this dependence \cite{SpinneyFord12a}. For
the systems under consideration in this paper, the distinction has
no meaning since for simplicity we assume the rates to be time-independent,
though for clarity the labels are retained. The definition of entropy
change according to Eq. (\ref{eq:5}) is compatible with other treatments
based on master equations \cite{Schnakenberg76,Harris07}.

Fig. \ref{fig:Trajectories-in-phase} indicates two further paths
that are related to $\boldsymbol{x}$, the probabilities of which
are to be established using a set of transition rates $T^{{\rm ad}}$
known as adjoint dynamics \cite{hatanosasa,Jarpathintegral}, and
given by
\begin{equation}
T^{{\rm ad}}(i^{\prime},m^{\prime}\vert i,m)=T(i,m\vert i^{\prime},m^{\prime})P^{{\rm st}}(i^{\prime},m^{\prime})/P^{{\rm st}}(i,m),\label{eq:4a}
\end{equation}
where the $P^{{\rm st}}(i,m)$ are the phase space probabilities in
the stationary state, which is by definition the same for both forward
and adjoint dynamics. The adjoint dynamics are designed to generate
a flux of probability in the stationary state that is opposite to
that which arises from the forward dynamics. Path $\boldsymbol{x}^{R}\equiv\left\{ \boldsymbol{x}_{N},\cdots,\boldsymbol{x}_{0}\right\} $
is the explicit reverse sequence of points, without velocity inversion,
and $\boldsymbol{x}^{T}\equiv\left\{ \boldsymbol{\varepsilon}\boldsymbol{x}_{0},\cdots,\boldsymbol{\varepsilon}\boldsymbol{x}_{N}\right\} $
is the forward sequence of spatial points but with inverted velocities.
These paths allow us to define path-dependent entropy change-like
quantities
\begin{equation}
\Delta\mathcal{S}_{1}(\boldsymbol{x})=\ln\left(P^{{\rm F}}(\boldsymbol{x})/P^{{\rm ad,R}}(\boldsymbol{x}^{R})\right),\label{eq:6}
\end{equation}
where $P^{{\rm ad,R}}(\boldsymbol{x}^{R})$ is the probability of
generating the path $\boldsymbol{x}^{R}$ under a reversed protocol
of time dependence of the transition rates of adjoint dynamics, and
\begin{equation}
\Delta\mathcal{S}_{2}(\boldsymbol{x})=\ln\left(P^{{\rm F}}(\boldsymbol{x})/P^{{\rm ad,F}}(\boldsymbol{x}^{T})\right),\label{eq:7}
\end{equation}
in a similar notation, where $P^{{\rm ad,F}}(\boldsymbol{x}^{T})$
is the probability of generating the path $\boldsymbol{x}^{T}$ under
adjoint dynamics with a forward protocol of time dependence. These
quantities each satisfy an integral fluctuation relation $\langle\exp(-\Delta\mathcal{S}_{{\rm tot},1,2})\rangle^{F}=1$,
where the brackets and superscript denote an average over all paths
generated by the forward dynamics from a given distribution of initial
coordinates, which implies that the expectation values of $\Delta\mathcal{S}_{{\rm tot}}$,
$\Delta\mathcal{S}_{1}$ and $\Delta\mathcal{S}_{2}$ over the forward
dynamics are never negative \cite{hatanosasa,IFThousekeeping,Jarpathintegral,adiabaticnonadiabatic0,adiabaticnonadiabatic1}.
Furthermore, for paths taken by a system when in a stationary state,
$\Delta\mathcal{S}_{1}$ is identically zero: it then consists of
a sum of changes to system and environmental or medium entropies that
cancel \cite{adiabaticnonadiabatic1,SpinneyFord12a}.

The forms taken by $\Delta\mathcal{S}_{1}$, $\Delta\mathcal{S}_{2}$
and $\Delta\mathcal{S}_{{\rm tot}}$ imply that we can represent the
total entropy change as a sum of three components: $\Delta\mathcal{S}_{{\rm tot}}=\Delta\mathcal{S}_{1}+\Delta\mathcal{S}_{2}+\Delta\mathcal{S}_{3}$.
In contrast to the other terms, the path average of the contribution
\begin{equation}
\Delta\mathcal{S}_{3}(\boldsymbol{x})=\ln\frac{P^{{\rm ad,R}}(\boldsymbol{x}^{R})P^{{\rm ad,F}}(\boldsymbol{x}^{T})}{P^{{\rm R}}(\boldsymbol{x}^{\dagger})P^{{\rm F}}(\boldsymbol{x})}\label{eq:8}
\end{equation}
does not have a fixed sign: however the average of this quantity in
a stationary state is zero.

Note that the total entropy production in stochastic thermodynamics
may also be divided into the components
\begin{equation}
\Delta\mathcal{S}_{{\rm tot}}=\Delta\mathcal{S}_{{\rm sys}}+\Delta\mathcal{S}_{{\rm med}},\label{eq:8a}
\end{equation}
corresponding to changes in the entropies of the system and surrounding
medium, respectively \cite{seifertoriginal,seifertprinciples}, and
furthermore that the latter can be divided into `housekeeping' and
`excess' contributions $\Delta\mathcal{S}_{{\rm med}}=\Delta\mathcal{S}_{{\rm hk}}+\Delta\mathcal{S}_{{\rm ex}}$
according to the descriptive scheme for nonequilibrium processes proposed
by Oono and Paniconi \cite{oono}. The correspondence with the present
division of entropy change is $\Delta\mathcal{S}_{1}=\Delta\mathcal{S}_{{\rm sys}}+\Delta\mathcal{S}_{{\rm ex}}$
and $\Delta\mathcal{S}_{2}+\Delta\mathcal{S}_{{\rm 3}}=\Delta\mathcal{S}_{{\rm hk}}$.
The terms $\Delta\mathcal{S}_{2}$ and $\Delta\mathcal{S}_{3}$ may
be designated the generalised and transient housekeeping contributions
to entropy production, respectively \cite{SpinneyFord12a}. $\Delta\mathcal{S}_{1}$
and $\Delta\mathcal{S}_{2}$ can also be mapped, respectively, onto
the nonadiabatic and adiabatic entropy productions of Esposito and
Van den Broeck \cite{adiabaticnonadiabatic0}, but only for systems
described by coordinates that are even under time reversal, or for
which the stationary state is symmetric in odd variables.

We can write $\Delta\mathcal{S}_{1}$ in terms of the time dependent
phase space probabilities $P(i,m,t)$, more compactly denoted $P(\boldsymbol{x}_{k},t)$,
that satisfy the master equation (\ref{eq:1a}) and an appropriate
initial condition, together with the corresponding distribution in
the stationary state, $P^{{\rm st}}(\boldsymbol{x}_{k})$, as follows:

\begin{equation}
\!\Delta\mathcal{S}_{{\rm 1}}\!(\boldsymbol{x})\!=\!\!\sum_{j=0}^{N}\!\ln\!\frac{P(\boldsymbol{x}_{j},t_{j}\!)}{P(\boldsymbol{x}_{j},t_{j+1}\!)}+\!\sum_{j=0}^{N-1}\!\ln\!\frac{P(\boldsymbol{x}_{j},t_{j+1}\!)P^{{\rm st}}(\boldsymbol{x}_{j+1}\!)}{P(\boldsymbol{x}_{j+1},t_{j+1}\!)P^{{\rm st}}(\boldsymbol{x}_{j}\!)}.\label{eq:8b}
\end{equation}
It is useful to associate each term with features of the path. The
first sum represents contributions due to residence at $\boldsymbol{x}_{j}$
in the periods $t_{j}\le t\le t_{j+1}$ and the second relates to
transitions from $\boldsymbol{x}_{j}$ to $\boldsymbol{x}_{j+1}$
at time $t_{j+1}$. Similarly, we can write $\Delta\mathcal{S}_{2}$
in terms of the transition rates $T$ and the stationary probabilities
$P^{{\rm st}}(\boldsymbol{x}_{k})$:
\begin{align}
 & \Delta\mathcal{S}_{{\rm 2}}(\boldsymbol{x})=\sum_{j=0}^{N}\left(t_{j+1}-t_{j}\right)\left(T(\boldsymbol{x}_{j}|\boldsymbol{x}_{j})-T(\boldsymbol{\varepsilon}\boldsymbol{x}_{j}|\boldsymbol{\varepsilon}\boldsymbol{x}_{j})\right)\nonumber \\
 & \qquad\qquad+\sum_{j=0}^{N-1}\ln\!\frac{P^{{\rm st}}(\boldsymbol{\varepsilon}\boldsymbol{x}_{j})}{P^{{\rm st}}(\boldsymbol{\varepsilon}\boldsymbol{x}_{j+1})}\frac{T(\boldsymbol{x}_{j+1}|\boldsymbol{x}_{j})}{T(\boldsymbol{\varepsilon}\boldsymbol{x}_{j}|\boldsymbol{\varepsilon}\boldsymbol{x}_{j+1})}.\label{eq:9}
\end{align}
 Again, the first sum may be viewed as contributions due to residence
at point $\boldsymbol{x}_{j}$ for a period $t_{j+1}-t_{j}$, and
the second sum may be associated with the transitions. Finally, we
write
\begin{equation}
\Delta\mathcal{S}_{3}(\boldsymbol{x})=\sum_{j=0}^{N-1}\ln\!\frac{P^{{\rm st}}(\boldsymbol{x}_{j})P^{{\rm st}}(\boldsymbol{\varepsilon}\boldsymbol{x}_{j+1})}{P^{{\rm st}}(\boldsymbol{x}_{j+1})P^{{\rm st}}(\boldsymbol{\varepsilon}\boldsymbol{x}_{j})},\label{eq:10}
\end{equation}
which is a sum of contributions associated with the transitions. Clearly
$\Delta\mathcal{S}_{3}$ vanishes if none of the phase space coordinates
change sign under time reversal ($\boldsymbol{\varepsilon}\boldsymbol{x}_{j}=\boldsymbol{x}_{j}$)
or if the stationary state is symmetric in odd variables ($P^{{\rm st}}(\boldsymbol{x}_{j})=P^{{\rm st}}(\boldsymbol{\varepsilon}\boldsymbol{x}_{j})$).

The entropy production associated with a specific path is straightforward
to compute from the elementary contributions arising from residence
and transition. The path $\boldsymbol{x}$ in Fig. \ref{fig:Trajectories-in-phase},
for example, consists of residence at four points and three transitions
between them: both $\Delta\mathcal{S}_{1}$ and $\Delta\mathcal{S}_{2}$
are therefore sums of seven terms, while $\Delta\mathcal{S}_{3}$
is a sum of only three.

The meaning of the above formalism is explored next using simple stochastic
systems that involve velocity as well as spatial coordinates on a
discrete grid. The systems are easier to analyse than those described
by dynamics in continuous coordinates, and so we can readily calculate
contributions to entropy production, both for individual paths and
when averaged over all possible paths, and understand better their
distinct physical origins, properties and meaning.

\section{Asymmetric telegraph process}

The telegraph or Kac process is a well known example of stochastic
dynamics on a phase space involving both position and velocity \cite{Kac74}.
For our purposes, it is best understood as an elaboration, in the
following way, of a standard asymmetric random walk in one dimension.
At times separated by a fixed interval of length $\Delta t$, a particle
chooses the direction of its next step to be to the right or left
with probabilities $(c+a)/2c$ and $(c-a)/2c$, respectively, with
$c\ge0$ and $-c\le a\le c$. It then spends the subsequent interval
$\Delta t$ performing the move at a constant speed. The position
and direction of motion of the particle are recorded halfway through
each timestep. At the next recording, the particle will have either
returned to its previous position but with the opposite velocity,
having reversed its direction of motion in between, or will have moved
to an adjacent spatial position without having changed its direction.
These events may be modelled using a master equation of the form given
in Eq. (\ref{eq:1a}) evolving the probabilities $P(i,\pm,t)$ that
the particle should assume position $X_{i}$ and velocity $\pm$ at
time $t$, and with the rates $T(X_{i+1}+\vert X_{i}+)=c+a$, $T(X_{i}-\vert X_{i}+)=c-a$,
$T(X_{i-1}-\vert X_{i}-)=c-a$ and $T(X_{i}+\vert X_{i}-)=c+a$, namely
\begin{equation}
\begin{aligned}\frac{dP(i,+,t)}{dt}= & (c+a)\left(P(i-1,+,t)+P(i,-,t)\right)\\
 & -2cP(i,+,t)\\
\frac{dP(i,-,t)}{dt}= & (c-a)\left(P(i+1,-,t)+P(i,+,t)\right)\\
 & -2cP(i,-,t).
\end{aligned}
\label{eq:10a}
\end{equation}
 We assume spatially periodic boundary conditions, such that the stationary
phase space probabilities are $P^{{\rm st}}(i,\pm)=(c\pm a)/(2cL)$.
Notice that $a/c$ plays the role of the dimensionless nonequilibrium
constraint parameter in this example: if it were zero, then the stationary
state would be symmetric in velocity and detailed balance in the sense
of $T(X_{i+1}+\vert X_{i}+)P^{{\rm st}}(i,+)=T(X_{i}-\vert X_{i+1}-)P^{{\rm st}}(i+1,-)$
would hold.

\begin{figure}
\includegraphics[width=1\columnwidth]{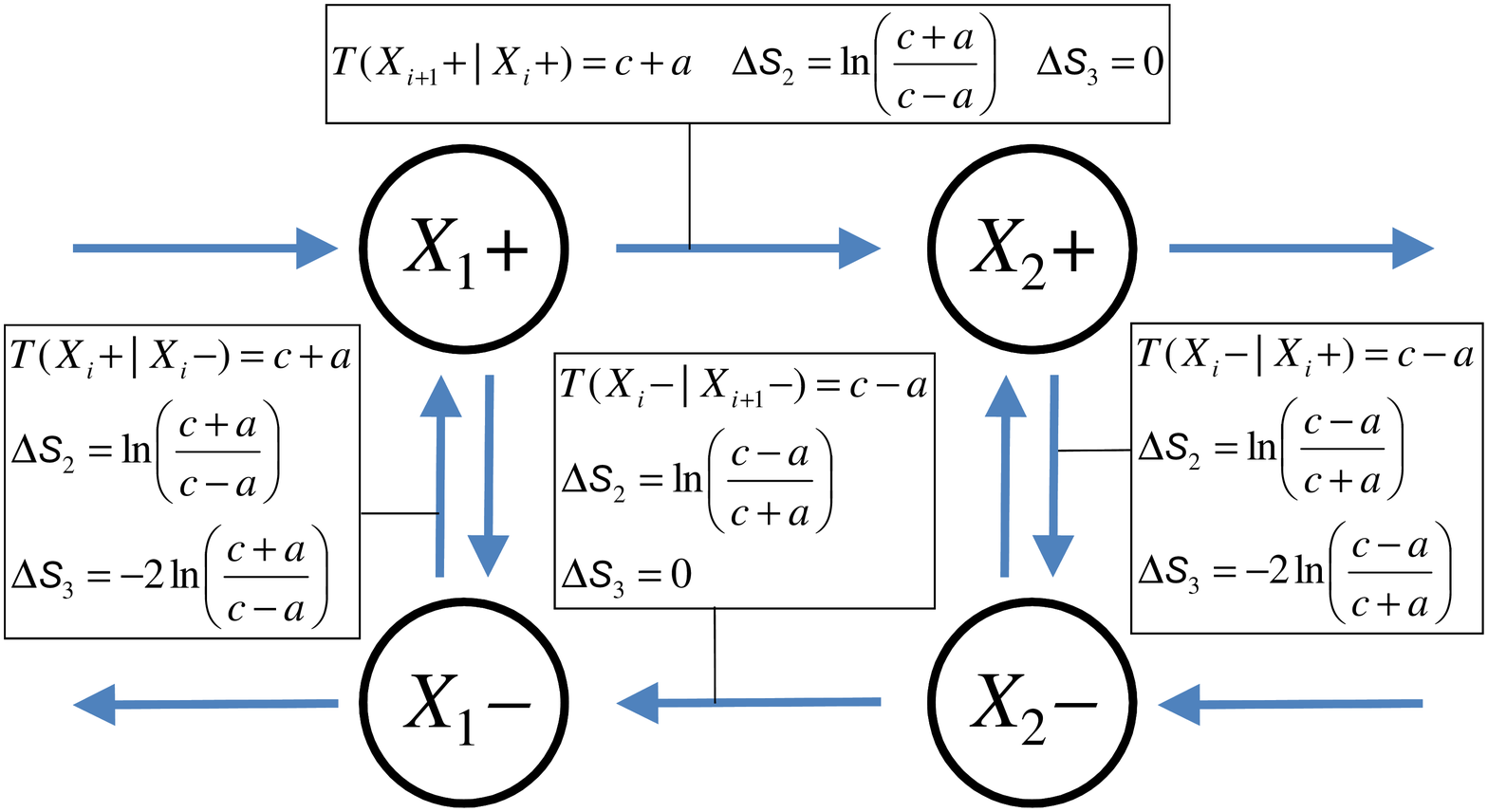}

\caption{Contributions to path-dependent housekeeping-type entropy production
in an asymmetric telegraph or Kac process. A particle follows a random
walk in one spatial dimension with periodic boundary conditions, on
a phase space defined in terms of position $X_{i}$ and direction
of motion $\pm$. The transition rates $T$ in the appropriate master
equation are shown, together with values of $\Delta\mathcal{S}_{2}$
and $\Delta\mathcal{S}_{3}$ specific to each step. \label{fig:telegraph}}
\end{figure}

We have all the ingredients needed to compute the entropy production.
A section of the phase space and the contributions to the housekeeping-type
entropy changes $\Delta\mathcal{S}_{2}$ and $\Delta\mathcal{S}_{3}$
made by each of the four distinct transitions are shown in Fig. \ref{fig:telegraph}.
There are no contributions to $\Delta\mathcal{S}_{2}$ from residence
at each site in this example, since $T(X_{i}+\vert X_{i}+)=T(X_{i}-\vert X_{i}-)=-2c$.
The contributions to $\Delta\mathcal{S}_{1}$ depend on the $P(i,\pm,t)$
and are not shown in the diagram for reasons of clarity.

For a given path followed by a particle across the phase space, it
is clear from the diagram that the contribution to $\Delta\mathcal{S}_{3}$
depends solely on whether the velocity has changed sign. The distribution
of $\Delta\mathcal{S}_{3}$ over an ensemble of paths generated by
the dynamics is therefore trimodal on the values $\pm2\ln[(c+a)/(c-a)]$
and zero. In contrast, the distribution of $\Delta\mathcal{S}_{2}$
depends on the net spatial displacement from the initial position,
as well as whether the velocity has changed sign,  and is broader,
but still discrete in units of $\ln[(c+a)/(c-a)]$.

\begin{figure}
\includegraphics[width=1\columnwidth]{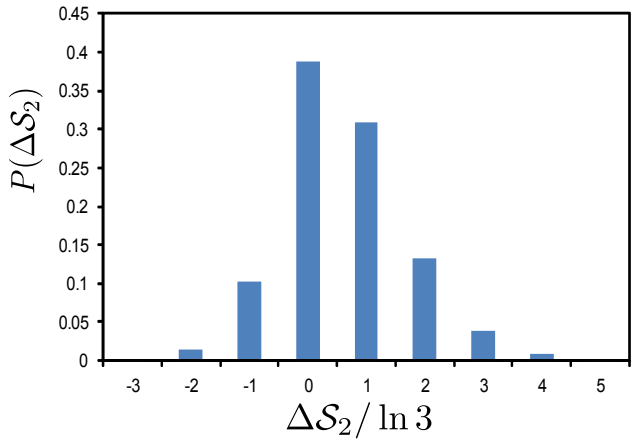}

\caption{Probability distribution of values of $\Delta\mathcal{S}_{2}$, in
units of $\ln3$, for test case A: a stochastic process on the phase
space in Fig. \ref{fig:telegraph} starting from a specific site with
positive velocity, and for parameters $c=2$, $a=1$ and $t=0.3$.
\label{fig:ds2hist}}

\end{figure}

To illustrate this, let us introduce test case A: evolution from an
initial distribution with $P(K,+,0)=1$, for a specified $K$, and
$P(i,\pm,0)=0$ for all other points. From consideration of the changes
to $\Delta\mathcal{S}_{2}$ associated with each transition in Fig.
\ref{fig:telegraph}, we can deduce that if the particle is found
at point $X_{K+k}+$ at time $t$, having reached it by any path with
a winding number around the periodic boundaries of zero, then the
accumulated $\Delta\mathcal{S}_{2}$ is equal to $k\ln[(c+a)/(c-a)]$.
Similarly, if the particle has reached point $X_{K+k}-$ the accumulated
$\Delta\mathcal{S}_{2}$ is $(k-1)\ln[(c+a)/(c-a)]$. The probabilities
$P(K+k,\pm,t)$ are straightforward to obtain from the master equations
(\ref{eq:10a}), and for early times are dominated by contributions
with zero winding number. As an illustration, we solve Eqs. (\ref{eq:10a})
numerically with $c=2$, $a=1$, $L=10$ and elapsed time $t=0.3$,
for the given initial condition, and provide a histogram of the resulting
distribution $P(\Delta\mathcal{S}_{2})$ in Fig. \ref{fig:ds2hist}.
The distribution can be used to verify the expected integral fluctuation
relation $\sum\exp(-\Delta\mathcal{S}_{2})P(\Delta\mathcal{S}_{2})=1$.
Furthermore, it satisfies a detailed fluctuation relation $P(\Delta\mathcal{S}_{2})=\exp(\Delta\mathcal{S}_{2})P(-\Delta\mathcal{S}_{2})$.
In contrast, the distribution of $\Delta\mathcal{S}_{3}$ for test
case A is $P(\Delta\mathcal{S}_{3}=0)=0.825$, $P(\Delta\mathcal{S}_{3}=2\ln3)=0.175$,
corresponding to the summed probabilities of positive and negative
velocity states, respectively, for these conditions, and does not
satisfy either fluctuation relation.

\begin{figure}
\includegraphics[width=1\columnwidth]{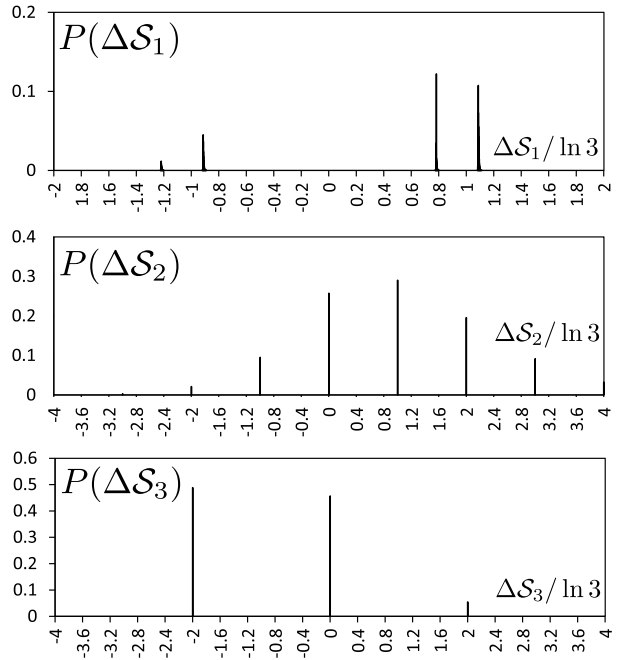}

\caption{Probability distributions of the entropy production $\Delta\mathcal{S}_{1}$,
$\Delta\mathcal{S}_{2}$ and $\Delta\mathcal{S}_{3}$ accumulated
in the period $0\le t\le0.5$ for test case B: evolution away from
a stationary state of the asymmetric telegraph process for $t<0$
with $a=-1$ and $c=2$, induced by setting $a=1$ for $t\ge0$. The
entropy production is given in units of $\ln3$. Note that the distribution
for $\Delta\mathcal{S}_{1}$ is continuous whilst those for $\Delta\mathcal{S}_{2}$
and $\Delta\mathcal{S}_{3}$ are discrete. \label{fig:ds}}

\end{figure}

An illustration of the stochastic evolution of $\Delta\mathcal{S}_{1}$
may be obtained by considering a transition between stationary states
brought about by a change in sign of the parameter $a$ (from negative
to positive) at $t=0$. The master equations are readily solved to
give
\begin{equation}
P(i,\pm,t)=\frac{c\pm a}{2cL}\mp\frac{a}{cL}\exp(-2ct),\label{eq:41a}
\end{equation}
and using these we can identify the contributions to $\Delta\mathcal{S}_{1}$
that arise from residence and transitions between phase space points:
\begin{equation}
\begin{aligned} & \Delta\mathcal{S}_{1}(X_{i}+\to X_{i}-)=\ln\frac{\left(c-a\right)\left(c+a\left(1-2e^{-2ct}\right)\right)}{\left(c+a\right)\left(c-a\left(1-2e^{-2ct}\right)\right)}\\
 & \Delta\mathcal{S}_{1}(X_{i}-\to X_{i}+)=-\Delta\mathcal{S}_{1}(X_{i}+\to X_{i}-)\\
 & \Delta\mathcal{S}_{1}(X_{i}+\to X_{i}+,\Delta t)=\ln\frac{\left(c+a\left(1-2e^{-2ct}\right)\right)}{\left(c+a\left(1-2e^{-2c(t+\Delta t)}\right)\right)}\\
 & \Delta\mathcal{S}_{1}(X_{i}-\to X_{i}-,\Delta t)=\ln\frac{\left(c-a\left(1-2e^{-2ct}\right)\right)}{\left(c-a\left(1-2e^{-2c(t+\Delta t)}\right)\right)},
\end{aligned}
\label{eq:14f}
\end{equation}
where $\Delta t$ is the residence period. There are no contributions
for transitions without a change in direction. These supplement the
assignments given in Fig. \ref{fig:telegraph} and using paths generated
by Monte Carlo simulation, can be employed to obtain probability distributions
$P(\Delta\mathcal{S}_{1})$, $P(\Delta\mathcal{S}_{2})$ and $P(\Delta\mathcal{S}_{3})$
associated with such an evolution away from a stationary state. We
explore this by introducing test case B with $c=2$, $a=-1$ for $t<0$
and $a=1$ for $t\ge0$, and illustrate the entropy generating behaviour
in Fig. \ref{fig:ds}. Probability distributions are determined for
elapsed time $t=0.5$ and can be used to verify that $P(\Delta\mathcal{S}_{1})$
and $P(\Delta\mathcal{S}_{2})$ satisfy an integral fluctuation relation.
Furthermore it may be shown that $P(\Delta\mathcal{S}_{2})$ satisfies
a detailed fluctuation relation.

Next we consider the mean value of each component of entropy production.
Clearly, if $a\ne0$ there is a nonequilibrium stationary state with
an overall probability current directed spatially to right or left
depending on the sign of $a$. The stationary state is then asymmetric
in the velocity coordinate and breaks detailed balance. The mean rate
of entropy production $d\langle\Delta\mathcal{S}_{2}\rangle/dt$ is
a sum of transition-specific contributions weighted by the phase space
probabilities and transition rates, and with reference to Fig. \ref{fig:telegraph}
it is given by
\begin{equation}
\begin{aligned} & \frac{d\langle\Delta\mathcal{S}_{2}\rangle}{dt}=\sum_{i=1}^{L}P(i,+,t)\left((c+a)\ln\left(\frac{c+a}{c-a}\right)\right.\\
 & \left.+(c-a)\ln\left(\frac{c-a}{c+a}\right)\right)\\
 & +\sum_{i=1}^{L}P(i,-,t)\left((c-a)\ln\left(\frac{c-a}{c+a}\right)\right.\\
 & \left.+(c+a)\ln\left(\frac{c+a}{c-a}\right)\right)=2a\ln\left(\frac{c+a}{c-a}\right).
\end{aligned}
\label{eq:40}
\end{equation}
Note that there are no terms associated with residence. The form of
this expression indicates that $d\langle\Delta\mathcal{S}_{2}\rangle/dt$
is never negative, consistent with the integral fluctuation relation.
It is also time independent, whether the system is in the stationary
state or relaxing towards it, and its magnitude depends on the condition
for the breakage of detailed balance $a\ne0$. We deduce that $\langle\Delta\mathcal{S}_{2}\rangle=2a\ln[(c+a)/(c-a)]t$
and this result may be confirmed explicitly for the distribution in
Fig. \ref{fig:ds2hist} for test case A, and the distribution in Fig.
\ref{fig:ds} for test case B.

\begin{figure}
\includegraphics[width=1\columnwidth]{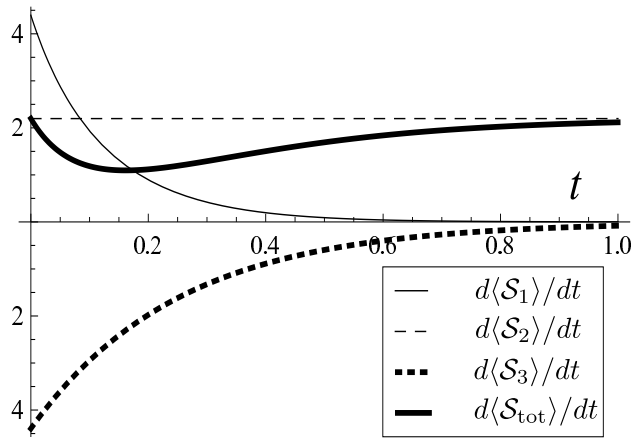}

\caption{Mean rates of entropy production against time for test case B: a transition
between nonequilibrium stationary states of the asymmetric telegraph
process. The initial state is characterised by $c=2$ and $a=-1$
and a total mean rate of entropy production equal to $2\ln3$. At
$t=0$ the parameter $a$ changes instantaneously to $+1$, and the
mean rate of change of $\Delta\mathcal{S}_{1}$ (thin solid line),
$\Delta\mathcal{S}_{2}$ (dashed) and $\Delta\mathcal{S}_{3}$ (dotted)
evolve as shown. The mean rate of total entropy production (thick
solid line) decreases temporarily but is eventually restored to its
initial value. \label{fig:telegraph entropy}}

\end{figure}

The rate $d\langle\Delta\mathcal{S}_{3}\rangle/dt$ may be computed
similarly:
\begin{equation}
\begin{aligned}\frac{d\langle\Delta\mathcal{S}_{3}\rangle}{dt}=\sum_{i=1}^{L}P(i,+,t)\left(-2(c-a)\ln\left(\frac{c-a}{c+a}\right)\right)\\
+\sum_{i=1}^{L}P(i,-,t)\left(-2(c+a)\ln\left(\frac{c+a}{c-a}\right)\right)\\
=2\left(c\sum_{i=1}^{L}\left(P(i,+,t)-P(i,-,t)\right)-a\right)\ln\left(\frac{c+a}{c-a}\right).
\end{aligned}
\label{eq:41}
\end{equation}
This expression has no bounds on its sign, but it vanishes in the
stationary state when $P(i,\pm,t)=P^{{\rm st}}(i,\pm)=(c\pm a)/(2cL)$.
It can be integrated for the conditions of test case A to obtain $\langle\Delta\mathcal{S}_{3}\rangle=0.35\ln3$
at $t=0.3$, which is compatible with the distribution of $\Delta\mathcal{S}_{3}$
reported earlier. And for the evolution away from a stationary state
caused by switching the sign of $a$, we can insert Eq. (\ref{eq:41a})
into Eq. (\ref{eq:41}) to obtain
\begin{equation}
\frac{d\langle\Delta\mathcal{S}_{3}\rangle}{dt}=-4a\exp(-2ct)\ln\left(\frac{c+a}{c-a}\right).\label{eq:42a}
\end{equation}
Finally, we consider the increment $\delta\langle\Delta\mathcal{S}_{1}\rangle$
over an increment in time $\delta t$ using Eq. (\ref{eq:8b}):
\begin{equation}
\begin{aligned} & \!\!\delta\langle\Delta\mathcal{S}_{1}\rangle\!=\!\sum_{i=1}^{L}\! P(i,+,t)(c-a)\delta t\Delta\mathcal{S}_{1}(X_{i}+\to X_{i}-)\\
 & +\!\sum_{i=1}^{L}\! P(i,-,t)(c+a)\delta t\Delta\mathcal{S}_{1}(X_{i}-\to X_{i}+)\\
 & +\!\sum_{i=1}^{L}\! P(i,+,t)(1-2c\delta t)\Delta\mathcal{S}_{1}(X_{i}+\to X_{i}+,\delta t)\\
 & +\!\sum_{i=1}^{L}\! P(i,-,t)(1-2c\delta t)\Delta\mathcal{S}_{1}(X_{i}-\to X_{i}-,\delta t).
\end{aligned}
\label{eq:14c}
\end{equation}
Using the $P(i,\pm,t)$ from Eq. (\ref{eq:41a}) for the transition
between stationary states, and considering terms to first order in
$\delta t$, we obtain $d\langle\Delta\mathcal{S}_{1}\rangle/dt$
for such situations:
\begin{equation}
\!\frac{d\langle\Delta\mathcal{S}_{1}\rangle}{dt}\!=\!2a\exp(-2ct)\ln\!\left(\!\frac{1+2a\exp(-2ct)/(c-a)}{1-2a\exp(-2ct)/(c+a)}\!\right).\label{eq:14e}
\end{equation}
The three mean contributions together with their sum $d\langle\Delta\mathcal{S}_{{\rm tot}}\rangle/dt$
are shown in Fig. \ref{fig:telegraph entropy} for test case B. The
two transient contributions $d\langle\Delta\mathcal{S}_{1}\rangle/dt$
and $d\langle\Delta\mathcal{S}_{3}\rangle/dt$ cancel initially, but
their differing time dependence apparent in Eqs. (\ref{eq:14e}) and
(\ref{eq:42a}) gives rise to a short-lived reduction in the mean
rate of total entropy production as the transition proceeds. The system
reorganises itself until the initial mean rate of entropy production
is restored. The integrals of the curves between $t=0$ and $t=0.5$
correspond to the means of the distributions in Fig. \ref{fig:ds}.

It is instructive to contrast this behaviour with the mean entropy
production associated with the asymmetric random walk when described
in terms of spatial position alone. An appropriate master equation
for probabilities $P(i,t)$ on a spatial grid with periodic boundaries
can be constructed using transition rates to right and left of $(c+a)$
and $(c-a)$ respectively. The stationary probabilities are $P^{{\rm st}}(i)=L^{-1}$
and a transition between stationary states brought about by a change
in sign of $a$ does not disturb them. Hence $d\langle\Delta\mathcal{S}_{1}\rangle/dt$
and $d\langle\Delta\mathcal{S}_{3}\rangle/dt$ are zero and the total
mean rate of entropy production corresponds to $d\langle\Delta\mathcal{S}_{2}\rangle/dt$,
which arises from contributions to $\Delta\mathcal{S}_{2}$ of $\ln[(c+a)/(c-a)]$
for a step to the right, and of $\ln[(c-a)/(c+a)]$ for a step to
the left. The mean rate of total entropy production in a stationary
state is hence
\begin{equation}
\begin{aligned} & \!\!\frac{d\langle\Delta\mathcal{S}_{{\rm tot}}\rangle}{dt}=\frac{d\langle\Delta\mathcal{S}_{2}\rangle^{{\rm st}}}{dt}\!=\!\sum_{i=1}^{L}\!\left(\! P^{{\rm st}}(i)(c+a)\ln\!\left(\frac{c+a}{c-a}\right)\right.\\
 & \!\!\left.+P^{{\rm st}}(i)(c-a)\ln\left(\frac{c-a}{c+a}\right)\right)=2a\ln\left(\frac{c+a}{c-a}\right)
\end{aligned}
\label{eq:42-1}
\end{equation}
as before and is independent of time. Clearly the transition between
stationary states is not now accompanied by a change in the mean rate
of total entropy production. The reorganisation of probability with
respect to the velocity coordinate is neglected at this level of description
of the dynamics, and the time dependence in the mean rate of total
entropy production seen in the more detailed description in Fig. \ref{fig:telegraph entropy}
cannot be appreciated. The example serves to demonstrate that entropy
production depends upon the chosen level of coarse graining: it is
after all a measure of the irreversibility of a dynamical system,
and perceived irreversibility will depend on the detail employed in
modelling the dynamics.

\section{Diffusion and barrier crossing}

\begin{figure}
\includegraphics[width=1\columnwidth]{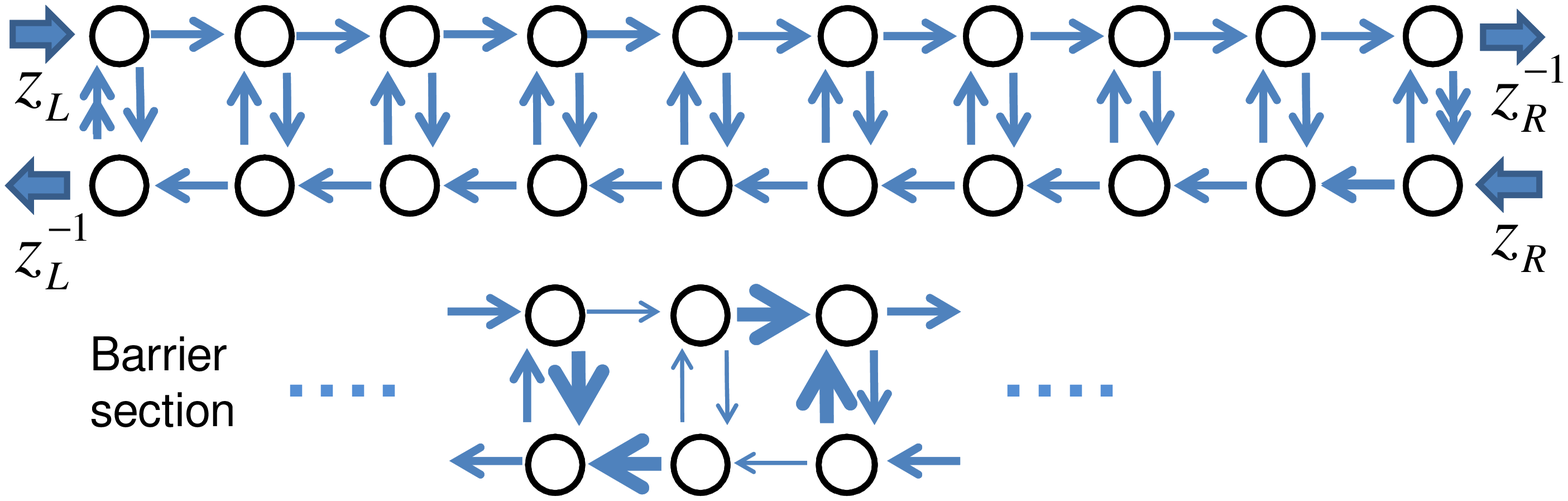}

\caption{A simple model of the dynamics of particle flow driven by a chemical
potential gradient, and its consequent entropy production. The upper
part of the diagram shows a phase space consisting of ten spatial
points each associated with two velocities, with transitions between
them indicated by various arrows described in the text. Particles
are injected and removed from the left and right hand sites at specific
rates. In the lower part of the diagram, a section of the phase space
with transition rates that represent a barrier to the flow is illustrated.
\label{fig:diffusion}}

\end{figure}

We elaborate the telegraph process now to illustrate the generation
of entropy associated with the diffusion of a particle across the
spatial extent of a system driven by a chemical potential gradient.
We consider the asymmetric random walk of a particle, as represented
by the telegraph process, but with $a=0$ for now, and for a system
with reflective boundaries. The latter condition implies that the
probability of a change in direction is unity when a particle is incident
upon a boundary, such that we should modify the transition rates in
those circumstances. This is illustrated in the upper part of Fig.
\ref{fig:diffusion} for a system with $L=10$ spatial points. Transitions
occurring at a rate $c$ are indicated with standard size single-headed
arrows. The double-headed arrows at positions $i=1$ and 10 represent
reflections at the boundaries and correspond to transition rates of
$2c$. In the absence of further processes, it is clear from counting
the rates of gain and loss at each point that the stationary phase
space probabilities for such a system are uniform both in position
and velocity: an equilibrium state corresponding to zero mean entropy
production.

Now consider transfers of a particle into and out of the system. Physically,
we can imagine the insertion and removal taking place at the spatial
boundaries to the left and right of the diagram. In order to include
the situation where the physical system is empty, we create an additional
phase space point available to the particle, but without spatial or
velocity coordinates. We allow the insertion of a particle from this
point into physical phase space points corresponding to inwardly directed
motion at the extreme left and right hand positions, as shown by block
arrows in Fig. \ref{fig:diffusion}. The rates of insertion are $z_{L}$
and $z_{R}$, respectively, and such rates appear in the relevant
master equations multiplied by the probability $P_{E}(t)$ that the
system should be empty. To balance these insertions, we include removals
from the extreme left and right hand positions, this time from the
phase space point corresponding to outwardly directed motion, again
shown by block arrows in Fig. \ref{fig:diffusion}. The rates of removal
are $z_{L}^{-1}$ and $z_{R}^{-1}$, respectively, such that an equilibrium
state can be established at $z_{L}=z_{R}=z$, with uniform stationary
probabilities defined by $P^{{\rm st}}(i,m)=P_{F}^{e}/(2L)$. The
probability $P_{F}^{e}$ that there is a particle somewhere in the
physical phase space when at equilibrium, divided by the equilibrium
probability $P_{E}^{e}$ that the system is empty, is then controlled
by $z$ according to $P_{F}^{e}/P_{E}^{e}=2Lz^{2}$. The insertion
and removal rates can be related to the exponential of the chemical
potential of a local particle bath and the approach resembles grand
canonical Monte Carlo. Thus if we were to consider the stochastic
dynamics under conditions $z_{L}>z_{R}$, we would expect the particle
to follow paths involving transfers into and out of the physical system,
producing a mean flux of probability from left to right across the
physical phase space, and a positive mean rate of entropy production.
The ratio $(z_{L}-z_{R})/z_{R}$ plays the role of the dimensionless
nonequilibrium constraint parameter for this case.

\begin{figure}
\includegraphics[width=1\columnwidth]{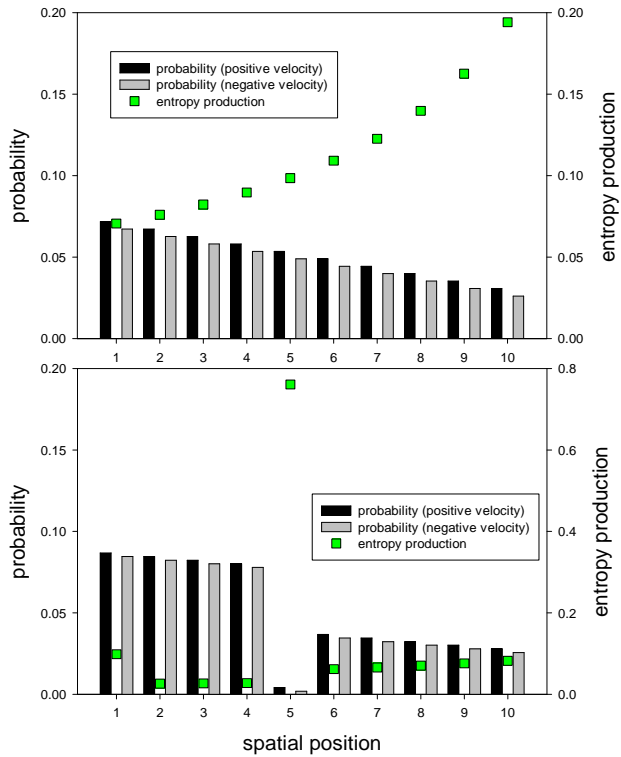}

\caption{The left and right hand columns at each spatial position represent
the stationary probabilities for each velocity in the phase space
illustrated in Fig. \ref{fig:diffusion}. The upper plot corresponds
to dynamical conditions $c=1$, $z_{L}=2$ and $z_{R}=1$, whilst
the lower plot includes the barrier section around position 5 with
$a=-0.9$. The square symbols denote the path-dependent contribution
$\Delta\mathcal{S}_{2}$ associated with a step to the right from
each spatial position (right hand axis). Passage across the barrier
is characterised by a relatively large local value of $\Delta\mathcal{S}_{2}$.
\label{fig:chemical-potential}}

\end{figure}

In order to explore this in detail, we consider the nonequilibrium
stationary state corresponding to the rates $c=1$, $z_{L}=2$ and
$z_{R}=1$. The stationary probabilities $P^{{\rm st}}(i,\pm)$ are
shown as vertical bars in the upper part of Fig. \ref{fig:chemical-potential}
together with points representing the increments in $\Delta\mathcal{S}_{2}$
associated with steps to the right starting from position $i$. Profiles
of path-dependent entropy increments associated with steps to the
left, and with reversals of direction, may also be generated but are
not shown. The increasing path-dependent contributions to entropy
production as a function of position are a consequence of the constant
probability current to the right, and the linear decrease in the stationary
probabilities. The situation is analogous to a stationary state of
particle diffusion between a source at high chemical potential on
the left and a sink at low chemical potential on the right.

This behaviour is modified if the transition rates are altered in
the fashion indicated in the lower part of Fig. \ref{fig:diffusion}.
The thicker and thinner arrows represent transition rates $c-a$ and
$c+a$ respectively, with $a<0$, and we can imagine the replacement
of the central part of the transition rate scheme by such a `barrier
section'. The replacement creates local external forces that impede
the particle flow between source and sink, and distort the stationary
probability distribution across the phase space. Nevertheless, the
probability distribution with $z_{L}=z_{R}$ even in the presence
of the barrier remains symmetric in the velocity coordinate and is
an equilibrium state. For a nonequilibrium stationary state with $z_{L}=2$,
$z_{R}=1$, $c=1$ and $a=-0.9$ (in the barrier section) the profile
of probabilities across phase space and the contributions to $\Delta\mathcal{S}_{2}$
associated with each spatial step to the right are illustrated in
the lower part of Fig. \ref{fig:chemical-potential}. The interpretation
is that the specific passage of the particle across the barrier is
associated with a relatively large contribution to entropy production.
Rare moves in a direction favoured by the prevailing thermodynamic
forces are more irreversible than commonplace ones: this is an intuitively
useful viewpoint, though we recognise it to be somewhat tautological
\cite{shargel}. The history of entropy production associated with
a stochastic path taken by a particle on this phase space will contain
periods during which the entropy fluctuates up and down, as the particle
explores the phase space region to the left of the barrier, which
are terminated by relatively large positive spikes in entropy production
when the particle crosses the barrier.

\begin{figure}
\includegraphics[width=1\columnwidth]{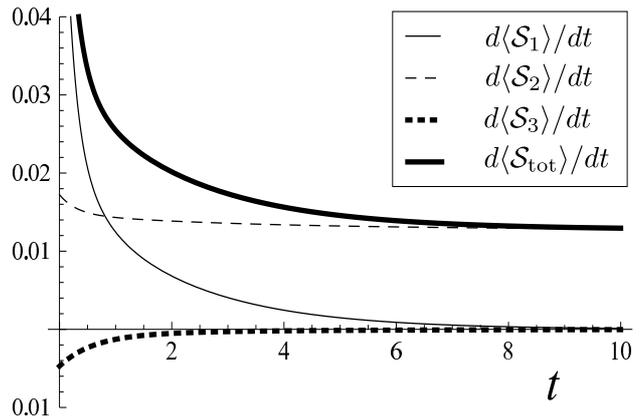}

\caption{Evolution of the mean rates of change of $\Delta\mathcal{S}_{1}$
(thin solid line), $\Delta\mathcal{S}_{2}$ (dashed), $\Delta\mathcal{S}_{3}$
(dotted) and their sum (thick solid line) corresponding to the transition
brought about by the change in $z_{L}$ from 1 to 2 at $t=0$ for
the system illustrated in the upper part of Fig. \ref{fig:diffusion}:
$z_{R}=1$ throughout so the mean rates for $t\le0$ are all zero.
\label{fig:chem-pot-transient}}

\end{figure}

The mean rates of change of the various components of entropy production
for the system without a barrier and with $c=1$, for a process starting
in equilibrium at $z_{L}=z_{R}=1$ for $t<0$ and driven out of equilibrium
by an instantaneous switch to $z_{L}=2$ for $t\ge0$, are shown in
Fig. \ref{fig:chem-pot-transient}. The system responds to the implied
change in chemical potential of the left hand source with mean entropy
production in all three components. For $t\le0$ the mean rate of
production of entropy is zero for all components, and so there is
a discontinuity in the rate of production when the process begins.
Once again, this pattern of mean entropy production differs with respect
to a model of the system that involves only the spatial phase space
points: in particular there would be no contribution $\Delta\mathcal{S}_{3}$
in the latter case. At such a coarser level of description the reorganisation
of the probability distribution over the velocity coordinate of phase
space would not be perceived, and consequently the assessment of the
irreversibility of the process would not be the same.

\section{dynamics on a ring with $M$ discrete velocities }

\begin{figure}
\includegraphics[width=1\columnwidth]{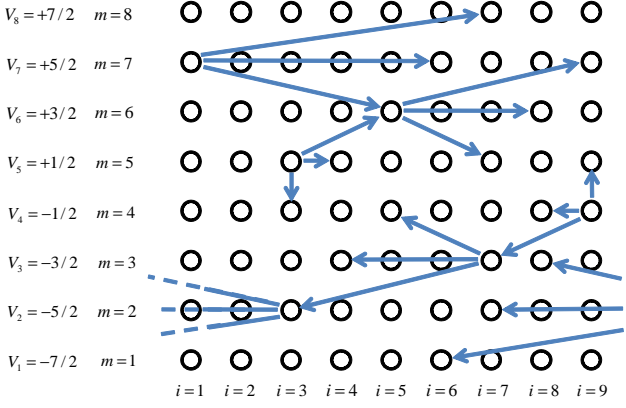}

\caption{A phase space of discrete spatial (labelled $i$) and velocity (labelled
$m$) coordinates with a selection of allowed moves of a particle
as described in Eqs. (\ref{eq:1-1}). Periodic boundary conditions
apply spatially.\label{fig:A-phase-space}}
\end{figure}

We now consider a phase space composed of discrete spatial positions
$X_{i}$ with periodic boundary conditions, as before, but now with
an extended set of available velocities $V_{m}=(m-1)-(M-1)/2$, where
$m=1,\cdots,M$ with $M$ an even integer. We define non-zero transition
rates to be
\begin{equation}
\begin{aligned} & T(i+\Delta i_{m}^{0},m|i,m)=C\\
 & T(i+\Delta i_{m}^{+},m+1|i,m)=\left(M-m\right)A/(M-1)\\
 & T(i+\Delta i_{m}^{-},m-1|i,m)=(m-1)B/(M-1),
\end{aligned}
\label{eq:1-1}
\end{equation}
where $A$, $B$ and $C$ are positive. The spatial displacements
$\Delta i_{m}^{0,\pm}$ are defined by
\begin{equation}
\Delta i_{m}^{0}=2(m-1)-(M-1),\;\Delta i_{m}^{\pm}=\Delta i_{m}^{0}\pm1,\label{eq:3-1}
\end{equation}
and some examples of transitions are illustrated in Fig. \ref{fig:A-phase-space}.
The dynamics represent changes of velocity according to the familiar
Ehrenfest model \cite{Ehrenfest07}, such that in the limit of large
$M$ the stationary velocity distribution approximates to a gaussian
envelope with a mean determined by $A$ and $B$. The spatial displacements
correspond to propagation at either the velocity $V_{m}$, or an average
of $V_{m}$ and $V_{m\pm1}$ if there should be a change in velocity,
for a period $\Delta t=2$.

\begin{figure}[t]
\includegraphics[width=1\columnwidth]{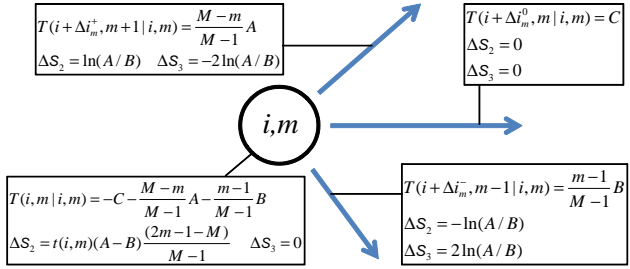}

\caption{An illustration of the four types of behaviour of a particle in the
phase space illustrated in Fig. \ref{fig:A-phase-space}, indicating
the associated transition rates $T$ and housekeeping-type contributions
$\Delta\mathcal{S}_{2}$ and $\Delta\mathcal{S}_{3}$ to entropy production.
The particle resides at phase space point $(X_{i},V_{m})\equiv(i,m)$
for a time $t(i,m)$ prior to a move described by one of the rates
in Eq. (\ref{eq:1-1}). The $\Delta\mathcal{S}_{1}$ take a form that
depends on the probabilities $P(i,m,t)$ and are specified in Eq.
(\ref{eq:8b}). \label{fig:A-particle-may}}
\end{figure}

In spite of some similarities in appearance, this model with $M=2$
is not equivalent to the asymmetric telegraph process discussed earlier:
the specification of the transition rates is different. In fact this
case is a generalisation of the two-velocity system considered previously
\cite{SpinneyFord12a}, and is a discrete version of a treatment of
driven Brownian motion of a particle on a ring \cite{SpinneyFord12b}.
The stationary probability distribution is
\begin{equation}
P^{{\rm st}}(i,m)=\frac{(M-1)!(A/B)^{m-1}}{L(M-m)!(m-1)!}\left(1+\frac{A}{B}\right)^{1-M},\label{eq:3}
\end{equation}
which is uniform over spatial coordinates, and is characterised by
a mean velocity equal to $(A-B)(M-1)/(2(A+B))$. The dimensionless
nonequilibrium constraint parameter is $(A-B)/B$: if this were zero,
the stationary distribution would be symmetric in velocity.

The transition rates and stationary probabilities allow us, as before,
to compute contributions to entropy production associated with the
detailed history of a path. Values of $\Delta\mathcal{S}{}_{2}$ and
$\Delta\mathcal{S}_{3}$ are illustrated for the model in Fig.~\ref{fig:A-particle-may}:
$\Delta\mathcal{S}_{1}$ is not included since it depends on $P(i,m,t)$
and is less compact in form. Note that there are now non-zero contributions
to $\Delta\mathcal{S}_{2}$ arising from residence at the phase space
points. These were absent in the telegraph process considered earlier,
as a consequence of the particular choice of transition rates.

Let us examine the means of entropy contributions $\Delta\mathcal{S}_{1}$,
$\Delta\mathcal{S}_{2}$ and $\Delta\mathcal{S}_{3}$ over a short
time interval $\delta t$. We average over the behaviour of the particle,
employing probabilities $1-T(\boldsymbol{x}_{j}|\boldsymbol{x}_{j})\delta t$
for residence over the period at $\boldsymbol{x}_{j}$, and probability
$T(\boldsymbol{x}_{j+1}|\boldsymbol{x}_{j})\delta t$ for a transition
from $\boldsymbol{x}_{j}$ to $\boldsymbol{x}_{j+1}$. We do not consider
multiple transitions during the interval since it is short. According
to Eq. (\ref{eq:8b}) the mean increment in $\Delta\mathcal{S}_{1}$
is
\begin{align}
 & \!\delta\langle\Delta\mathcal{S}_{1}\rangle\!=\nonumber \\
 & \!\!\sum_{i,m}P(i,m,t)\biggl(\frac{M-m}{M-1}A\delta t\ln\left(\!\frac{P(i,m,t)(M-m)A}{P(i,m+1,t)mB}\!\right)\nonumber \\
 & \!+\frac{m-1}{M-1}B\delta t\ln\!\left(\!\frac{P(i,m,t)(m-1)B}{P(i,m-1,t)(M-m+1)A}\!\right)\nonumber \\
 & \left.-\left(1-T(i,m|i,m)\delta t\right)\!\frac{d\ln P(i,m,t)}{dt}\!\delta t\right).\!\!\!\label{eq:10aa}
\end{align}
Note that the last term vanishes to order $\delta t$ due to normalisation.
We rearrange to get
\begin{align}
 & \!\!\frac{d\langle\Delta\mathcal{S}_{1}\rangle}{dt}\!=\!\frac{1}{M-1}\!\sum_{i}\sum_{m=1}^{M-1}\!\Bigl(P(i,m,t)(M-m)A\!\nonumber \\
 & \!\!-P(i,m+1,t)mB\Bigr)\ln\!\left(\!\frac{P(i,m,t)(M-m)A}{P(i,m+1,t)mB}\!\right),\!\!\label{eq:10b}
\end{align}
a form that is explicitly positive, or zero in a stationary state
when $P(i,m,t)=P^{{\rm st}}(i,m)$. Similarly
\begin{align}
 & \delta\langle\Delta\mathcal{S}_{2}\rangle=\sum_{i,m}P(i,m,t)\left((A-B)\frac{(2m-1-M)}{M-1}\right.\nonumber \\
 & \left.+\frac{M-m}{M-1}A\ln\!\left(\!\frac{A}{B}\!\right)-\frac{m-1}{M-1}B\ln\!\left(\!\frac{A}{B}\!\right)\!\right)\delta t,\label{eq:11}
\end{align}
for small $\delta t$, giving
\begin{align}
 & \!\frac{d\langle\Delta\mathcal{S}_{2}\rangle}{dt}\!=\!\sum_{i,m}\!\frac{P(i,m,t)}{M-1}\!\left(\! B(m-1)\!\left(\!\frac{A}{B}-1-\ln\!\left(\!\frac{A}{B}\!\right)\!\right)\right.\nonumber \\
 & \left.\qquad\qquad+A(M-m)\left(\!\frac{B}{A}-1-\ln\!\left(\!\frac{B}{A}\!\right)\!\right)\right),\label{eq:12}
\end{align}
which is also explicitly positive unless $A=B$, the condition for
a symmetric stationary state over velocities, in which case it vanishes.
Its mean in a stationary state is obtained by inserting Eq. (\ref{eq:3}):
\begin{equation}
\frac{d\langle\Delta\mathcal{S}_{2}\rangle^{{\rm st}}}{dt}=\frac{(A-B)^{2}}{A+B}.\label{eq:13}
\end{equation}
 Finally,
\begin{align}
 & \frac{d\langle\Delta\mathcal{S}_{3}\rangle}{dt}=\sum_{i,m}\frac{P(i,m,t)}{M-1}\left(-2(M-m)A\ln\left(A/B\right)\right.\nonumber \\
 & \left.\qquad\qquad+2(m-1)B\ln\left(A/B\right)\right),\label{eq:14}
\end{align}
which can take either sign, but which vanishes when the system is
in the stationary state or when $A=B$. These results with $M=2$
coincide with those obtained previously \cite{SpinneyFord12a}.

\begin{figure}
\includegraphics[width=1\columnwidth]{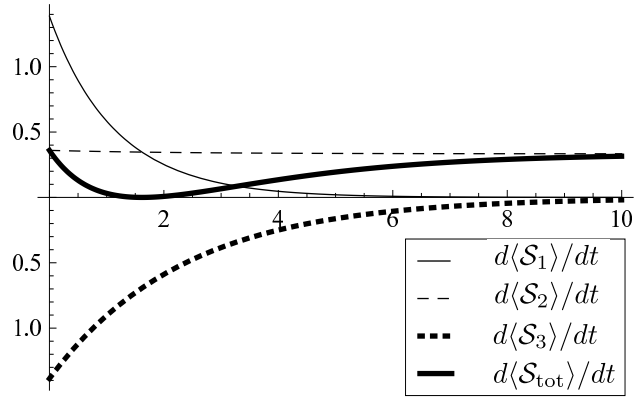}

\caption{Mean rates of entropy production against time for a transition between
stationary states for a particle on a ring with $M=8$ available velocities
and transition rules illustrated in Fig. \ref{fig:A-phase-space}
and Eq. (\ref{eq:1-1}). For $t<0$ the system is in a stationary
state parametrised by $A=1$ and $B=2$. For $t\ge0$, these parameter
values are swapped and the system makes a transition to a new stationary
state. \label{fig:extended}}

\end{figure}

The importance of the velocity coordinate for the deeper appreciation
of irreversibility in this system may be demonstrated by determining
the mean entropy production associated with a transition between stationary
states brought about by the instantaneous swapping of the values of
the parameters $A$ and $B$. From a perspective of a treatment in
a phase space of positions alone, there would be no relaxation of
the probability distribution: the only entropy production brought
about by the transition would arise from $\Delta\mathcal{S}_{2}$,
as was found in test case B for the telegraph process. However, according
to Eq. (\ref{eq:13}), which remains valid for this coarser grained
treatment, the mean rate of entropy production would not change. Only
from a perspective of the more detailed dynamics would there be additional
contributions $\Delta\mathcal{S}_{1}$ and $\Delta\mathcal{S}_{3}$,
arising from the relaxation in the probability distribution over velocities.
The latter behaviour is illustrated in Fig. \ref{fig:extended} for
an example with parameters $A=2$ and $B=1$ with $M=8$, $C$ being
arbitrary for this example. The total mean rate of entropy production
falls momentarily to zero during the transition, as is also seen in
the treatment of driven Brownian motion on a continuous phase space
reported elsewhere \cite{SpinneyFord12b}. The deviation from the
constant contribution $d\langle\Delta\mathcal{S}_{2}\rangle/dt$ is
due to the relaxational terms $d\langle\Delta\mathcal{S}_{1}\rangle/dt$
and $d\langle\Delta\mathcal{S}_{3}\rangle/dt$. Without a consideration
of velocity coordinates, the implication that the system behaves less
irreversibly, on average, during the transition between stationary
states would be missed. We conclude again that coarse-graining alters
the perception of irreversibility. It is worth noting, however, that
the behaviour seen in Fig. \ref{fig:extended} is similar that which
emerges from a continuum treatment of the same system \cite{SpinneyFord12b},
suggesting that dynamics on a coarse-grained, discrete representation
of a continuous phase space can still capture certain features of
the irreversibility.

\section{A simple model of thermal conduction}

\begin{figure}
\includegraphics[width=0.9\columnwidth]{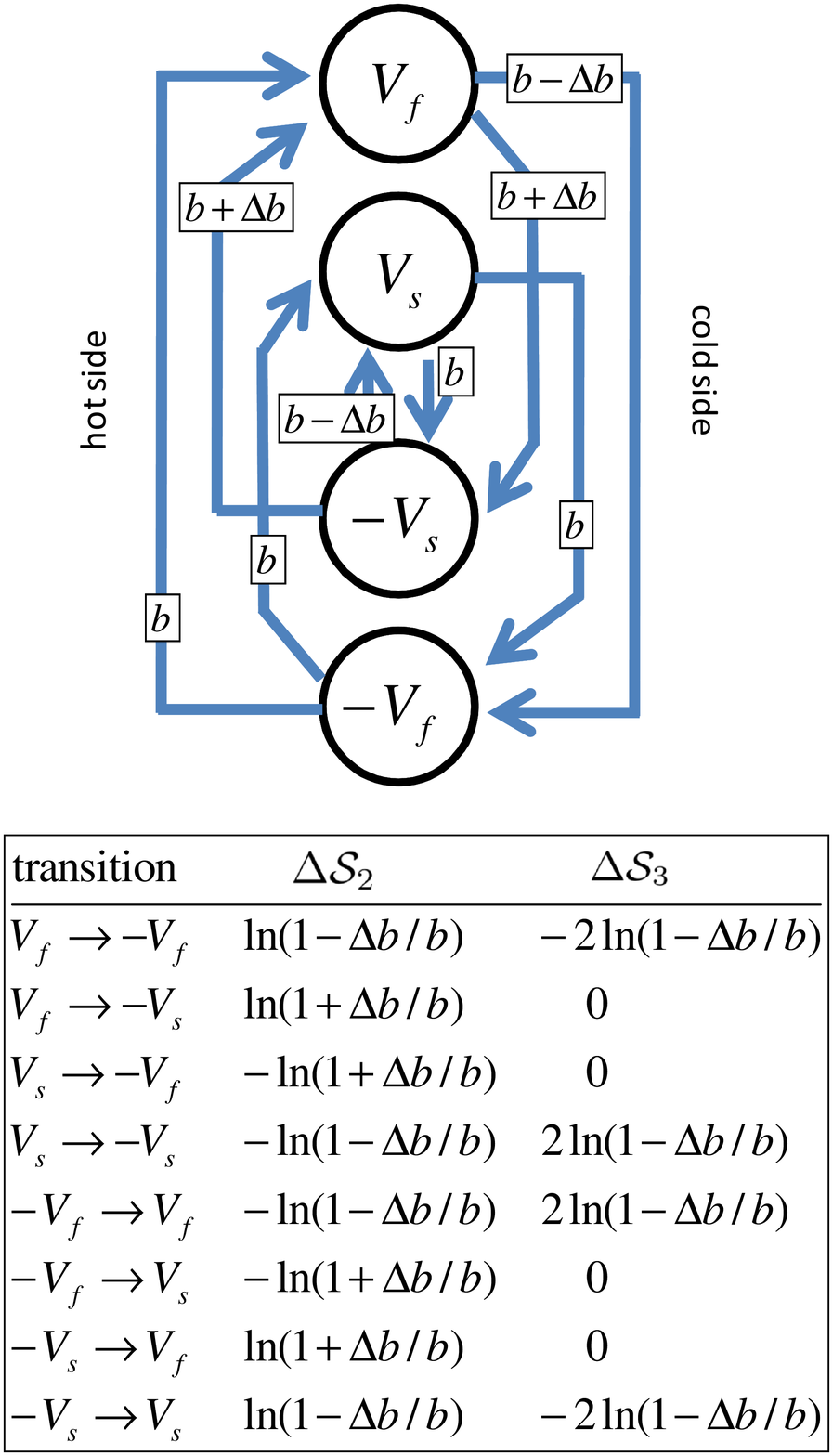}

\caption{A simple model of thermal conduction involving one spatial point and
four values of velocity. A particle is reflected from hot and cold
boundaries and the consequent transitions between fast and slow velocity
coordinates in each direction are represented by arrows with specific
rates shown. The contributions $\Delta\mathcal{S}_{2}$ and $\Delta\mathcal{S}_{3}$
to entropy production associated with each transition are given as
a function of the thermal gradient parameter $\Delta b$. \label{fig:thermal}}

\end{figure}

Our final example system is illustrated in Fig. \ref{fig:thermal}.
It is one of the simplest phase space dynamical schemes that can represent
thermal conduction and the associated production of entropy. The phase
space consists of a single spatial position, with four possible velocities:
fast (suffix $f$) and slow (suffix $s$) in each direction. The transitions
between the points correspond to the reflection of a particle from
boundaries situated to the left and right of the spatial position.
The left hand boundary has the property that slow arrivals are reflected
with probability $(b+\Delta b)/(2b)$ into the fast returning velocity
state and with probability $(b-\Delta b)/(2b)$ into the slow state,
with transition rate $b>0$ and $-b\le\Delta b\le b$. For positive
$\Delta b$ the bias towards the fast return velocity suggests we
can regard the left hand boundary as `hot'. For simplicity, the left
hand boundary is assumed to partition fast arrivals into fast and
slow returning states with equal probability. In contrast, the right
hand boundary reflects fast arrivals with probability $(b+\Delta b)/(2b)$
into the slow return state, and with probability $(b-\Delta b)/(2b)$
into the fast state. Slow arrivals at this boundary are partitioned
with equal probabilities into fast and slow returns. These rules allow
us to regard the right hand boundary as `cold'. More elaborate schemes
than these might be conceived, but the symmetries in the transitions
and associated rates, as illustrated in Fig. \ref{fig:thermal}, simplify
the analysis of entropy production (in particular, there are no contributions
to $\Delta\mathcal{S}_{2}$ from residence at the phase space points)
while allowing the model to capture something of the physics of thermal
conduction, with $\Delta b$ representing a thermal gradient across
the system and $\Delta b/b$ the dimensionless nonequilibrium constraint
parameter.

The contributions to $\Delta\mathcal{S}_{2}$ and $\Delta\mathcal{S}_{3}$
associated with the eight available transitions in the system are
listed in Fig. \ref{fig:thermal}. These are calculated on the basis
of the transition rates indicated in the diagram and the stationary
state probabilities $P^{{\rm st}}(V_{f})=P^{{\rm st}}(-V_{s})=b/(4b-2\Delta b)$
and $P^{{\rm st}}(V_{s})=P^{{\rm st}}(-V_{f})=(b-\Delta b)/(4b-2\Delta b)$.
Since $V_{f}>V_{s}$, this implies that the mean particle velocity
is directed towards the right for $\Delta b>0$. Clearly some paths
undertaken by the particle between the boundaries, such as $V_{f}\to-V_{f}\to V_{f}$,
give no net change in entropy, whilst those that involve a slowing
down at the cold boundary and speeding up at the hot boundary (for
example $V_{f}\to-V_{s}\to V_{f}$) produce a positive increment for
$\Delta b>0$. In contrast, the circuit $V_{s}\to-V_{f}\to V_{s}$
produces a negative increment, compatible with a distribution of both
positive and negative values of $\Delta\mathcal{S}_{2}$.

The mean rate of change of $\Delta\mathcal{S}_{2}$ in the stationary
state may be shown to be
\begin{align}
 & \frac{d\langle\Delta\mathcal{S}_{2}\rangle^{{\rm st}}}{dt}\!=\! P^{{\rm st}}(V_{f})\!\left((b-\Delta b)\ln\left(1-\frac{\Delta b}{b}\right)\right.\!\nonumber \\
 & \!\left.+(b+\Delta b)\ln\left(1+\frac{\Delta b}{b}\right)\right)\!+\! P^{{\rm st}}(V_{s})\!\left(\!-b\ln\left(1+\frac{\Delta b}{b}\right)\right.\!\!\nonumber \\
 & \!\left.-b\ln\left(1-\frac{\Delta b}{b}\right)\right)+\! P^{{\rm st}}(-V_{f})\!\left(-b\ln\left(1-\frac{\Delta b}{b}\right)\right.\!\nonumber \\
 & \left.-b\ln\left(1+\frac{\Delta b}{b}\right)\right)+\! P^{{\rm st}}(-V_{s})\!\left((b+\Delta b)\ln\left(1+\frac{\Delta b}{b}\right)\right.\!\nonumber \\
 & \left.+(b-\Delta b)\ln\left(1-\frac{\Delta b}{b}\right)\right),\label{eq:15a}
\end{align}
which reduces to
\begin{equation}
\frac{d\langle\Delta\mathcal{S}_{2}\rangle^{{\rm st}}}{dt}=\frac{2b\Delta b}{2b-\Delta b}\ln\left(1+\frac{\Delta b}{b}\right),\label{eq:15b-1}
\end{equation}
a form that is never negative for the physical range $-b\le\Delta b\le b$.
For $\Delta b=0$ we recover the equilibrium state where $d\langle\Delta\mathcal{S}_{2}\rangle^{{\rm st}}/dt$
vanishes and where the phase space probabilities are symmetric in
velocity.

The mean value of $\exp(-\Delta\mathcal{S}_{2})$ for an incremental
time period $\delta t$ is
\begin{align}
 & \!\langle\exp(-\Delta\mathcal{S}_{2})\rangle=\! P(V_{f},t)\!\left((b-\Delta b)\delta t\left(1-\frac{\Delta b}{b}\right)^{\!\!-1}\right.\!\nonumber \\
 & \!\left.+(b+\Delta b)\delta t\left(1+\frac{\Delta b}{b}\right)^{\!\!-1}\right)\!+\! P(V_{s},t)\!\left(\! b\delta t\left(1+\frac{\Delta b}{b}\right)\right.\!\!\nonumber \\
 & \!\left.+b\delta t\left(\!1-\frac{\Delta b}{b}\right)\!\right)\!+\! P(-V_{f},t)\!\left(b\delta t\left(1-\frac{\Delta b}{b}\right)\right.\!\nonumber \\
 & \left.+b\delta t\left(\!1+\frac{\Delta b}{b}\right)\!\right)\!+\! P(-V_{s},t)\!\left(\!(b+\Delta b)\delta t\left(\!1+\frac{\Delta b}{b}\right)^{\!\!-1}\right.\!\nonumber \\
 & \left.+(b-\Delta b)\delta t\left(1-\frac{\Delta b}{b}\right)^{-1}\right)+\left(P(V_{f},t)+P(V_{s},t)\right.\nonumber \\
 & \left.+P(-V_{f},t)+P(-V_{s},t)\right)\left(1-2b\delta t\right),\label{eq:16}
\end{align}
where the last term is made up of contributions arising from residence
at each point. This reduces to $\langle\exp(-\Delta\mathcal{S}_{2})\rangle=1$
for the period $\delta t$, a result that therefore also holds for
a finite time period by consideration of a sequence of incremental
periods under the Markovian dynamics.

By a similar analysis, the mean rate of change of $\Delta\mathcal{S}_{3}$
in the stationary state is
\begin{align}
 & \frac{d\langle\Delta\mathcal{S}_{3}\rangle^{{\rm st}}}{dt}=-2P^{{\rm st}}(V_{f})(b-\Delta b)\ln\left(1-\frac{\Delta b}{b}\right)\nonumber \\
 & +\!2P^{{\rm st}}(V_{s})b\ln\left(1-\frac{\Delta b}{b}\right)+2P^{{\rm st}}(-V_{f})b\ln\left(1-\frac{\Delta b}{b}\right)\nonumber \\
 & -2P^{{\rm st}}(-V_{s})(b-\Delta b)\ln\left(1-\frac{\Delta b}{b}\right)=0,\label{eq:15b}
\end{align}
vanishing as expected.

\begin{figure}[t]
\includegraphics[width=1\columnwidth]{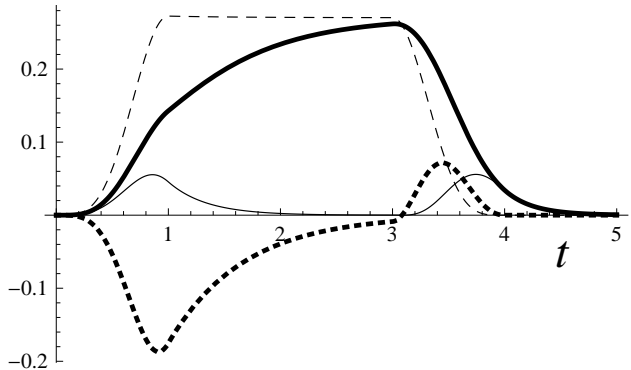}

\caption{The evolution of the mean rates of entropy production $d\langle\Delta\mathcal{S}_{1}\rangle/dt$
(thin solid line), $d\langle\Delta\mathcal{S}_{2}\rangle/dt$ (dashed),
$d\langle\Delta\mathcal{S}_{3}\rangle/dt$ (dotted) and their sum
$d\langle\Delta\mathcal{S}_{{\rm tot}}\rangle/dt$ (thick solid line)
for the system shown in Fig. \ref{fig:thermal} driven by $\Delta b(t)$
specified in Eq. (\ref{eq:17}). $d\langle\Delta\mathcal{S}_{3}\rangle/dt$
goes negative in response to the increase in $\Delta b$, and is largely
positive when $\Delta b$ decreases. All the other mean rates are
non-negative throughout since the distributions of those entropy increments
satisfy integral fluctuation relations. \label{fig:thermalmeans}}
\end{figure}

Transient departures from equilibrium can be investigated using a
numerical solution of the relevant master equation. As an example,
consider the imposition and removal of a thermal gradient brought
about by the specification
\begin{equation}
\Delta b(t)=\begin{cases}
\frac{1}{4}\left(1-\cos(\pi t)\right) & 0\le t\le1\\
\frac{1}{2} & 1\le t\le3\\
\frac{1}{4}\left(1+\cos(\pi(t-3))\right) & 3\le t\le4
\end{cases}\label{eq:17}
\end{equation}
with $\Delta b=0$ for $t<0$ and $t>4$, and $b=1$ throughout. The
mean rates $d\langle\Delta\mathcal{S}_{1}\rangle/dt$, $d\langle\Delta\mathcal{S}_{2}\rangle/dt$
and $d\langle\Delta\mathcal{S}_{3}\rangle/dt$, together with their
sum, are shown in Fig. \ref{fig:thermalmeans} over the course of
such a process. Several now-familiar features are apparent. The mean
rates of change of $\Delta\mathcal{S}_{1}$, $\Delta\mathcal{S}_{2}$
and $\Delta\mathcal{S}_{{\rm tot}}$, quantities that each satisfy
integral fluctuation relations, are never negative, whilst $d\langle\Delta\mathcal{S}_{3}\rangle/dt$
can take either sign. $d\langle\Delta\mathcal{S}_{2}\rangle/dt$ and
$d\langle\Delta\mathcal{S}_{3}\rangle/dt$ are non-zero only when
detailed balance is broken through the condition $\Delta b\ne0$.
The evolution of $d\langle\Delta\mathcal{S}_{2}\rangle/dt$ mirrors
the time dependence of $\Delta b(t)$ and both $d\langle\Delta\mathcal{S}_{1}\rangle/dt$
and $d\langle\Delta\mathcal{S}_{3}\rangle/dt$ are transient contributions
that tend to die away when $\Delta b$ is constant. Of the transient
terms, the production rate $d\langle\Delta\mathcal{S}_{1}\rangle/dt$
continues to evolve after $\Delta b(t)$ has gone to zero, whilst
$d\langle\Delta\mathcal{S}_{3}\rangle/dt$ vanishes in these circumstances.
The pattern of the mean rate of total entropy production $d\langle\Delta\mathcal{S}_{{\rm tot}}\rangle/dt$
provides a characterisation of the thermal conduction and its irreversibility.
It emerges with a delay in response to the time dependent thermal
gradient, which seems physically intuitive.

The contribution $d\langle\Delta\mathcal{S}_{3}\rangle/dt$ plays
an important part in establishing this pattern, and it relies on the
presence of odd velocity coordinates in the phase space. In fact it
would be impossible to describe the entropy production associated
with thermal conduction without considering the velocity coordinates
of a system: heat flow is the conveyance of kinetic energy. This has
already been established through an analysis of the process using
continuum stochastic dynamics \cite{SpinneyFord12b}. A heat current
with associated entropy production cannot be established in a phase
space of one spatial dimension with reflective boundaries and so dynamics
on a phase space of spatial points cannot capture the irreversibility
of heat conduction. This is the clearest of our demonstrations that
a failure to take account of dynamics in full phase space can lead
to the neglect of an important mechanism of entropy production and
alter the apparent irreversibility of the process in question.

\section{Conclusions}

We have investigated several simple examples where entropy production
can be associated with specific, stochastically generated paths taken
by a particle on a discrete full phase space. We have shown how it
is formed from three components, each with particular statistical
properties and a specific relationship to the underlying physical
origins of irreversibility. The component $\Delta\mathcal{S}_{1}$
is associated with relaxation towards a stationary state. $\Delta\mathcal{S}_{2}$
is associated with the breakage of detailed balance brought about
by the dynamical rules, and with entropy production in nonequilibrium
stationary states. $\Delta\mathcal{S}_{3}$ is also associated with
the breakage of detailed balance but only exists for dynamics that
include coordinates that change sign (are `odd') under time reversal
\cite{SpinneyFord12a,SpinneyFord12b}. A further condition for its
existence is that the probability distribution for the stationary
state under the prevailing conditions has to be asymmetric in an odd
coordinate. The mean of $\Delta\mathcal{S}_{3}$ is zero in a stationary
state and so, like $\Delta\mathcal{S}_{1}$, it is associated with
relaxation. Thus $\Delta\mathcal{S}_{3}$ has its origin in the dissipative
mechanisms that are separately responsible for $\Delta\mathcal{S}_{1}$
and $\Delta\mathcal{S}_{2}$. The three components can be related
to the changes in system and medium entropy \cite{seifertoriginal}
as well as, in some circumstances, adiabatic and nonadiabatic entropy
production \cite{adiabaticnonadiabatic0}.

Specifically, we have considered stochastic particle dynamics on discrete
full phase spaces with transition rules that capture aspects of drift
and diffusion, barrier crossing, injection, removal and interaction
with a thermal gradient. The examples are necessarily simplified,
but they illustrate the properties of the various components of entropy
production summarised in the last paragraph, and furthermore demonstrate
that the distributions of $\Delta\mathcal{S}_{1}$ and $\Delta\mathcal{S}_{2}$
satisfy integral fluctuation relations. They also show that the inclusion
of additional detail in the specification of a system, or its opposite,
coarse-graining, can have an impact on the assessment of entropy production.
This is compatible with the notion that entropy is a representation
of the uncertainty in microscopic state, and that its production is
linked to the dynamics employed when modelling the evolution. By taking
into account greater levels of detail in a dynamical system, we potentially
alter our perception of the uncertainty in the microscopic state,
and the pattern of entropy production could be modified as a result.
At the deepest level, of course, when we neglect no detail of either
system or environment, there will be no stochasticity in the model
(we would then employ the `real' dynamics) and no consequent development
of uncertainty, and from a point of view of stochastic thermodynamics
there would be no production of entropy. Entropy change in this perspective
is subjective, and depends on choices made in the modelling.

We emphasise that these conclusions are based on an interpretation
of entropy change and the second law acquired in the context of stochastic
thermodynamics. The underlying dynamics have been taken to be stochastic
and to break time reversal symmetry. Entropy change is then associated
with dynamical irreversibility expressed in terms of path probabilities.
An interpretation based on deterministic and time-reversible dynamics
is also available, whereby entropy generation is associated with the
contraction of a continuous phase space and it is possible to derive
fluctuation relations and a second law \cite{Evans94,Gallavotti2}.

In closing, and from a perspective of stochastic thermodynamics, we
reflect briefly on the nature of entropy and its production. Time
reversal invariance and determinism imply that events in the future
as well as the past will be apparent to a being capable of perceiving
the current state of the universe in all its detail, a point made
by Laplace. Loschmidt used the same reasoning to identify the flaw
in Boltzmann's initial attempts to build a mechanical model of entropy
production in an isolated ideal gas: the model could not provide the
arrow of time. But Loschmidt is far from being the villain of this
piece of the history of science \cite{Cercignani98}. Consideration
of time reversal enriches the meaning of entropy production in models
that explicitly break this symmetry, in many cases through a representation
of the interactions between a system and its environment that reflects
a state of perception inferior to that of Laplace's being. Symmetry
of equations of motion under time reversal is either present or it
is not, but the practical consequence of its absence emerges on a
broader spectrum, and in stochastic thermodynamics, entropy production
is its measure. Some trajectories generated by a stochastic model
are hard to reverse (meaning the reverse trajectory is relatively
unlikely); some are easier. The second law in stochastic thermodynamics
is a statement about the statistical dominance of the hard cases.
\begin{acknowledgments}
RES acknowledges financial support from the UK Engineering and Physical
Sciences Research Council.
\end{acknowledgments}
%

\end{document}